\documentclass[12pt]{article}
\usepackage{epsfig}
\usepackage{amsmath} 
\usepackage{amssymb} 

\newcommand\pubnumber{KA-TP-1-2001\\UAB-FT-506}
\newcommand\pubdate{January 2001}
\newcommand\hepnumber{hep-ph/0101294}

\def\csumb{$^{a}$ Grup de F{\'\i}sica Te\`{o}rica and Institut de
  F{\'\i}sica d'Altes Energies, Universitat Aut\`{o}noma de Barcelona, E-08193, Bellaterra, Barcelona, Catalonia, Spain\\ 
$^{b}$ Institut f\"{u}r Theoretische Physik, 
Universit{\"a}t Karlsruhe, Kaiserstra{\ss}e 12, D-76128 Karlsruhe,
Germany.}
\def\supportsb{\footnote{Supported in part by CICYT under project No. AEN99-0766.}} 
\def\supportjs{\footnote{Supported in part by the Direcci{\'o} General de
  Recerca de la Generalitat de Catalunya and by CICYT under project No. AEN99-0766.}} 
\def\supportjg{\footnote{Supported by the European Union under contract No. 
HPMF-CT-1999-00150.}} 

\def\Title#1{\begin{center} {\Large\bf #1 } \end{center}}
\def\Author#1{\begin{center}{ \sc #1} \end{center}}
\def\Address#1{\begin{center}{ \it #1} \end{center}}

\newcommand\pubblock{\rightline{\begin{tabular}{l} \pubnumber\\
         \pubdate\\ \hepnumber \end{tabular}}}
\newenvironment{Abstract}{\begin{quotation}  }{\end{quotation}}
\newenvironment{Presented}{\begin{quotation} \begin{center} 
             Invited talk presented at the \end{center}
      \begin{center}\begin{large}}{\end{large}\end{center} \end{quotation}}
\def\Acknowledgments{\bigskip  \bigskip \begin{center}
          \large\bf Acknowledgments\end{center}}

\makeatletter
\def\section{\@startsection{section}{0}{\z@}{5.5ex plus .5ex minus
 1.5ex}{2.3ex plus .2ex}{\large\bf}}
\def\subsection{\@startsection{subsection}{1}{\z@}{3.5ex plus .5ex minus
 1.5ex}{1.3ex plus .2ex}{\normalsize\bf}}
\def\subsubsection{\@startsection{subsubsection}{2}{\z@}{-3.5ex plus
-1ex minus  -.2ex}{2.3ex plus .2ex}{\normalsize\sl}}

\renewcommand{\@makecaption}[2]{%
   \vskip 10pt
   \setbox\@tempboxa\hbox{\small #1: #2}
   \ifdim \wd\@tempboxa >\hsize     
       \small #1: #2\par          
     \else                        
       \hbox to\hsize{\hfil\box\@tempboxa\hfil}
   \fi}

 \def\citenum#1{{\def\@cite##1##2{##1}\cite{#1}}}
 
\newcount\@tempcntc
\def\@citex[#1]#2{\if@filesw\immediate\write\@auxout{\string\citation{#2}}\fi
  \@tempcnta\z@\@tempcntb\m@ne\def\@citea{}\@cite{\@for\@citeb:=#2\do
    {\@ifundefined
       {b@\@citeb}{\@citeo\@tempcntb\m@ne\@citea\def\@citea{,}{\bf ?}\@warning
       {Citation `\@citeb' on page \thepage \space undefined}}%
    {\setbox\z@\hbox{\global\@tempcntc0\csname b@\@citeb\endcsname\relax}%
     \ifnum\@tempcntc=\z@ \@citeo\@tempcntb\m@ne
       \@citea\def\@citea{,}\hbox{\csname b@\@citeb\endcsname}%
     \else
      \advance\@tempcntb\@ne
      \ifnum\@tempcntb=\@tempcntc
      \else\advance\@tempcntb\m@ne\@citeo
      \@tempcnta\@tempcntc\@tempcntb\@tempcntc\fi\fi}}\@citeo}{#1}}
\def\@citeo{\ifnum\@tempcnta>\@tempcntb\else\@citea\def\@citea{,}%
  \ifnum\@tempcnta=\@tempcntb\the\@tempcnta\else
  {\advance\@tempcnta\@ne\ifnum\@tempcnta=\@tempcntb \else\def\@citea{--}\fi
    \advance\@tempcnta\m@ne\the\@tempcnta\@citea\the\@tempcntb}\fi\fi}
\makeatother

\def\beq{\begin{equation}}
\def\eeq#1{\label{#1}\end{equation}}
\def\eeqn{\end{equation}}

\newenvironment{Eqnarray}%
   {\arraycolsep 0.14em\begin{eqnarray}}{\end{eqnarray}}
\def\beqa{\begin{Eqnarray}}
\def\eeqa#1{\label{#1}\end{Eqnarray}}
\def\eeqan{\end{Eqnarray}}

\let\bar=\overbar

\def\Dslash{\not{\hbox{\kern-4pt $D$}}}
\def\dslash{\not{\hbox{\kern-2pt $\del$}}}

\def\mt{m_t}

\def\msb{{\bar{\ssstyle M \kern -1pt S}}}

\def\lsim{\mathrel{\raise.3ex\hbox{$<$\kern-.75em\lower1ex\hbox{$\sim$}}}}
\def\gsim{\mathrel{\raise.3ex\hbox{$>$\kern-.75em\lower1ex\hbox{$\sim$}}}}

\newcommand{\mg}{\ensuremath{m_{\tilde g}}}
\newcommand{\mA}{\ensuremath{m_{A^0}}}
\newcommand{\tb}{\ensuremath{\tan\beta}}  
\newcommand{\stackm}{\stackrel{\scriptstyle <}{{ }_{\sim}}}  
\newcommand{\stackM}{\stackrel{\scriptstyle >}{{ }_{\sim}}}

\begin{document}
\begin{titlepage}
\pubblock

\vfill
\def\thefootnote{\fnsymbol{footnote}}
\Title{FCNC top quark decays \\ beyond the Standard Model}
\vfill
\Author{Santi B\'{e}jar$^{a}$\supportsb, Jaume Guasch$^{b}$\supportjg, 
\underline{Joan Sol\`{a}}$^{a}$\supportjs}
\Address{\csumb}
\vfill
\begin{Abstract}
\noindent 
Flavor Changing Neutral Current decays of the top quark within the strict
context of the Standard Model are 
known to be extremely rare. In fact, they are hopelessly undetectable at
the Tevatron, LHC and LC in any of their scheduled upgradings.
Therefore, if a few of these events eventually show up in the future we
will have certainly
discovered new physics. We argue that this could well be the case for
the LHC and the LC both within the Minimal Supersymmetric Standard Model
(MSSM) and in a general two-Higgs-doublet model (2HDM),
especially if we look for FCNC top quark decays into Higgs bosons.
\end{Abstract}
\vfill
\begin{Presented}
5th International Symposium on Radiative Corrections \\ 
(RADCOR--2000) \\[4pt]
Carmel CA, USA, 11--15 September, 2000
\end{Presented}
\vfill
\end{titlepage}
\def\thefootnote{\arabic{footnote}}
\setcounter{footnote}{0}
%

\section{Introduction} 
At the tree-level there are no Flavor Changing Neutral Current (FCNC)
processes in the Standard Model (SM), and at one-loop 
they are induced by charged-current interactions, which are GIM-suppressed. 
In particular, FCNC decays of the top quark into gauge bosons ($t\rightarrow 
c\,V$;$\;V\equiv\gamma,Z,g$) are very unlikely 
($BR(t\rightarrow c\,\gamma,\,Z)\sim10^{-13}$ and 
$BR(t\rightarrow c\,g)\sim10^{-11}$)~\cite{Eilam:1991zc}. 
These are much smaller than the FCNC rates of a typical 
low-energy meson decay, e.g.~$B(b\rightarrow s\,\gamma)\sim10^{-4}$. The 
reason is simple: for FCNC top quark decays in the SM, the loop amplitudes 
are controlled by down-type quarks, mainly by the bottom quark. Therefore, 
the scale of the loop amplitudes is set by $m_{b}^{2}$ and the partial 
widths are of order
\begin{equation} 
\Gamma(t\rightarrow 
V\,c)\sim|V_{tb}^{\ast}V_{bc}|^{2}\alpha\,G_{F}^{2}\,m_{t}\,m_{b}^{4}\,F
\sim|V_{bc}|^{2}\alpha_{em}^2 \alpha \,m_{t}\left( \frac{m_{b}}{M_{W}}\right) ^{4}\,F, 
\label{GammaFCNC} 
\end{equation} 
where $\alpha$ is $\alpha_{em}$ for $V=\gamma,Z$ and $\alpha_{s}$ for $V=g$. 
The factor $F\sim(1-m_{V}^{2}/m_{t}^{2})^{2}$ results, upon neglecting $m_{c}$, 
from phase space and polarization sums. The fourth power mass ratio, in
parenthesis in eq.~(\ref{GammaFCNC}), stems from the GIM mechanism and
is responsible for the ultralarge  
suppression beyond naive expectations based on pure dimensional analysis, 
power counting and CKM matrix elements. From that simple formula, the approximate orders of 
magnitude mentioned above ensue immediately. 
 
Even more dramatic is the situation with the top quark decay into the SM 
Higgs boson, $t\rightarrow c\,H_{SM}$:
$BR(t\rightarrow c\,H_{SM})\sim10^{-13}-10^{-15}$
$(m_{t}=175\,GeV;\;M_{Z}\leq M_{H}\leq2\;M_{W})$~\cite{Mele:1998ag}. 
This extremely tiny rate is 
far out of the range to be covered by any presently conceivable high 
luminosity machine. On the other hand, the highest FCNC top quark rate in 
the SM, namely that of the gluon channel $t\rightarrow c\,g$, is still $6$ 
orders of magnitude below the feasible experimental possibilities at the 
LHC. All in all the detection of FCNC decays of the top quark at visible 
levels (viz. $\,BR(t\rightarrow c\,X)>10^{-5}$) by the high luminosity 
colliders round the corner seems doomed to failure 
in the absence of new physics. 
Unfortunately, although the FCNC decay modes 
into electroweak gauge bosons $V_{ew}=\gamma,Z$ may be enhanced a few orders of 
magnitude, it proves to be insufficient to raise the meager SM rates 
mentioned before up to detectable limits, and this is true both in the 2HDM 
-- where $BR(t\rightarrow V_{ew}\,c)<10^{-6}$~\cite{Eilam:1991zc} -- and in the 
MSSM -- where $BR(t\rightarrow V_{ew}\,c)<10^{-7}$~\cite{Lopez:1997xv}.
In this respect it is a lucky 
fact that these bad news need not to apply to the gluon channel, which could 
be barely visible ($BR(t\rightarrow g\,c)\stackm10^{-5}$) both in the
MSSM~\cite{deDivitiis:1997sh,Guasch:1999jp} and in the general 2HDM~\cite{Eilam:1991zc}. But, most  
significant of all, they may not apply to the non-SM Higgs boson
channels $t\rightarrow(h^{0},H^{0},A^{0})+c$ either. As we shall show,  
these Higgs decay channels of the top quark could lie above the visible 
threshold for a parameter choice made in perfectly sound regions of 
parameter space. 
 
A systematic discussion of these ``gifted'' Higgs channels has been made
in Ref.~\cite{Guasch:1999jp} for the MSSM and more recently in
Ref.~\cite{Bejar:2000ub} for the general 2HDM.
Here we will present the results in the 2HDM and the MSSM, and make a
close comparison between them.
We believe that this study is 
necessary, not only to assess what are the chances to see traces of new 
physics in the new colliders but also to 
clear up the nature of the virtual effects; in particular to disentangle 
whether the origin of the hypothetically detected FCNC decays of the top 
quark is ultimately triggered by SUSY or by some alternative 
renormalizable extension of the SM such as the 2HDM or generalizations 
thereof. Of course the alleged signs of new physics could be searched for 
directly through particle tagging, if the new particles were not too heavy. 
However, even if accessible, the corresponding signatures could be far from 
transparent. In contrast, the indirect approach based on the FCNC processes 
has the advantage that one deals all the time with the dynamics of the top 
quark. Thus by studying potentially new features beyond the well-known SM 
properties of this quark one can hopefully uncover the existence of the 
underlying new interactions~\cite{TopPhys}. 

\section{Relevant fields and interactions}

We will mainly focus our interest on the loop induced FCNC decays
\begin{equation} 
t\rightarrow c\;h\;\;\;(h=h^{0},H^{0},A^{0})\,,  \label{Higgschannels} 
\end{equation} 
in which any of the three possible neutral Higgs bosons from a general 2HDM 
can be in the final state. However, as a reference we shall compare 
throughout our analysis the Higgs channels with the more conventional gluon 
channel $t\rightarrow c\;g$. 

Although other quarks could participate in the final state of these 
processes, their contribution is negligible.
The lowest order diagrams entering these decays are 
one-loop diagrams in which Higgs, quarks, gauge and Goldstone bosons -- in 
the Feynman gauge -- circulate around. 
In the MSSM also the SUSY-parters of the above fields,
namely squarks and charginos, circulate in the loops. In addition there 
exists the possibility that the  squark-squared-mass-matrix 
is not simultaneously diagonal to the quark-mass-matrix. In this latter
case there exist tree-level FCNC couplings in the interactions
quark-squark-gluino and quark-squark-neutralino. This possibility is not
unnatural. If one computes the evolution of the
squark-squared-mass-matrix using the Renormalization Group Equations,
assuming alignment at a certain scale (e.g. a supposed Unification
Scale), one finds that non-diagonal terms for the
squark-left--squark-left entries are
generated~\cite{Duncan:1983iq}. 
We have computed the MSSM decay
widths under two different approximations: in the first one we assume
alignment, and the only induced FCNC are generated through the charged
sector of the model with the same mixing matrix as in the SM -- the
CKM-matrix; in the second approach we give up the alignment hypothesis,
and assume a free -- though restricted by
experiment~\cite{Gabbiani:1996hi}-- squark-mass-matrix
and compute the SUSY-QCD induced FCNC partial decay widths, which are the
leading ones under this approximation.

Here we follow the standard notation~\cite{Hunter}, namely $h^{0},H^{0}$
are CP-even Higgs bosons 
and $A^{0}$ is a CP-odd one. 
When the quark mass matrices are diagonalized in non-minimal
extensions of the Higgs sector of the SM, the Yukawa couplings do not in
general become simultaneously diagonalized, so that one would expect
Higgs mediated FCNC's at the tree-level. These are of course unwanted,
since they would lead to large FCNC processes in light quark
phenomenology, which are stringently restricted by experiment. 
One has two canonical choices to get rid of them, called
Type~I 2HDM and Type~II 2HDM~\cite{Hunter}.
The Higgs sector
of the MSSM is that of a Type~II 2HDM, with restrictions between the
parameters due to the SUSY constraints.

When analyzing the 2HDM~I, II cases we will use      
the following set of free parameters:  
\begin{equation} 
(m_{h^{0}},m_{H^{0}},m_{A^{0}},m_{H^{\pm}},\tan\alpha,\tan\beta )\,, 
\label{freeparam} 
\end{equation} 
where $m_{H^{\pm}}$ is the mass of the charged Higgs companions
$H^{\pm}$, $\tan\alpha$ defines the mixing angle $\alpha$ in the
diagonalization of the  
CP-even sector, and $\tan\beta$ gives the mixing angle $\beta$ in the CP-odd 
sector. The latter is a key parameter in our analysis. It is given by the 
quotient of the vacuum expectation values (VEV's) of the two Higgs
doublets $\Phi_{2,1}$, viz. $\tan\beta=v_{2}/v_{1}$~\cite{Hunter}. 
The most general (Type~I or Type~II) 2HDM
Higgs potential subject  
to hermiticity, $SU(2)\times U(1)$ and gauge invariance 
involves six scalar operators with six free 
(real) coefficients $\lambda_{i}\,(i=1,...,6)$
and the two VEV's~\cite{Hunter}. 
We will furthermore assume that $\lambda _{5}=\lambda_{6}$ in the general 2HDM Higgs 
potential~\cite{Bejar:2000ub}. 
The alternative set~(\ref{freeparam}) is just a (more physical)
reformulation of this fact after diagonalization of the mass matrices
and imposing the aforementioned set of constraints. The constraints
imposed by SUSY reduce the number of free parameters in
eq.~(\ref{freeparam}) to two, which we take to be $(m_{A^0},\tb)$, since
the radiative corrections to the rest of parameters~(\ref{freeparam})
are large we make use of the one-loop expressions to compute
them~\cite{Dabelstein}\footnote{The two-loop corrections to the Higgs
  sector of the MSSM have 
  also recently become available~\cite{Carena:1996wu}.  Their effect,
  however, cannot significantly modify our one-loop results.}.

The two canonical types of 2HDM's only differ in the
couplings to fermions but they share the rest of Feynman rules. Of
particular relevance are the rules for the trilinear Higgs vertices in
the 2HDM case, which depend on the Higgs boson mass differences and can
be enhanced for large and small $\tan\beta$ -- see
Ref.~\cite{Bejar:2000ub}. In the MSSM, 
however, the mass differences are correlated and one can further simplify
their form to a combination of trigonometric functions of $\alpha$ and
$\beta$, using the relations between the
parameters~(\ref{freeparam}) -- see Ref.~\cite{Hunter}.  We refrain from giving here
the interaction
Lagrangian~\cite{Guasch:1999jp,Bejar:2000ub,Hunter,Coarasa:1996qa}.

Both in the generic 2HDM~II and in the MSSM, the Feynman rules 
for the lightest CP-even Higgs, $h^{0}$, go over to the SM Higgs boson ones 
in the limit $\sin (\beta -\alpha )\rightarrow 1$. In the particular case of 
the MSSM, this limit is equivalent to $m_{A^{0}}\rightarrow \infty
$. Moreover, in the MSSM one has
$m_{h^{0}}\stackm 135\,GeV$~\cite{Carena:1996wu}
whereas in the general Type~II model there is no 
upper bound on $m_{h^{0}}$, and by the same token the corresponding lower 
bound is considerably less stringent -- see Ref.~\cite{Bejar:2000ub}. 
 
Since we shall perform our calculation in the on-shell scheme, we understand 
that the physical inputs are given by the electromagnetic coupling and the 
physical masses of all the particles.
It should be clear that, as there are no 
tree-level FCNC decays of the top quark, there is no need to introduce 
counterterms for the physical inputs. In fact, the 
calculation is carried out in lowest order with respect to 
the effective $tch$ and $tcg$ couplings and so the sum of all the one-loop 
diagrams (as well as of certain subsets of them) should be finite in a 
renormalizable theory, and indeed it is. 

From the interaction Lagrangians and Feynman rules it is 
straightforward to compute the loop induced FCNC rates for the decays
(\ref{Higgschannels}) and 
$t\rightarrow c\;g$~\cite{Guasch:1999jp,Bejar:2000ub}.
We shall refrain from listing the  
lengthy analytical formulae.  The computation in the MSSM was 
reported in great detail in Ref.~\cite{Guasch:1999jp}, and the one in
the 2HDM~\cite{Bejar:2000ub} is very similar. Therefore, we
will limit  
ourselves to exhibit the final numerical results. The fiducial ratio on 
which we will apply our numerical computation is the following:  
\begin{equation} 
B^{j}(t\rightarrow h+c)=\frac{\Gamma^{j}(t\rightarrow h+c)}{\Gamma 
(t\rightarrow W^{+}+b)+\Gamma^{j}(t\rightarrow H^{+}+b)}\,\,, 
\label{fiducialH} 
\end{equation} 
for each Type $j=I,II$ of 2HDM and the MSSM and for each neutral Higgs
boson $h=h^{0}$, $H^{0}$, $A^{0}$. While this ratio is not the total
branching fraction, it is enough for most practical purposes and it is
useful in order to compare with 
previous results in the literature. We define the fiducial branching
ratio for $t\to g+c$ in a similar way.

We have performed a fully-fledged independent analytical and numerical 
calculation of $\Gamma^{j}(t\rightarrow g+c)$ at one-loop in the context of 
2HDM~I, II and the MSSM. Where there is overlapping, we have checked the 
numerical results of Ref.~\cite{Eilam:1991zc}\footnote{In
  Ref.~\cite{Eilam:1991zc} $B(t\rightarrow g+c)$ is defined without 
including the charged Higgs channel contribution in the fiducial
branching ratio. The agreement is achieved, however, only if it is included.}.

Charged Higgs bosons from Type~II models are subject to an indirect
bound from the experimental measurement by CLEO of the branching
fraction $BR(B\rightarrow X_{s}\,\gamma)$~\cite{Alam:1995aw}.  From the
various analysis in the literature one finds
$m_{H^{\pm}}>(165-200)\,GeV$ for virtually any
$\tan\beta\stackM1$~\cite{Borzumati:1998tg,Chankowski:1999ta}.
This bound does not apply to Type~I models. 
Therefore, in principle the top quark decay~$t\rightarrow H^{+}+b$ is
still possible in 2HDM~I; but also in 2HDM~II, if $m_{H^{\pm}} $  
lies near the lowest end of the previous bound, and in this case that decay 
can contribute to the denominator of
eq.~(\ref{fiducialH}). In SUSY models this limit does
not apply provided $\mu A_t<0$ -- see e.g.~\cite{Ciuchini:1998xy}. 

\section{$t\to c h$ and $t\to c g$ in the MSSM} 
\begin{figure}[t]
\begin{center}
\begin{tabular}{cc}
\resizebox{5cm}{!}{\includegraphics{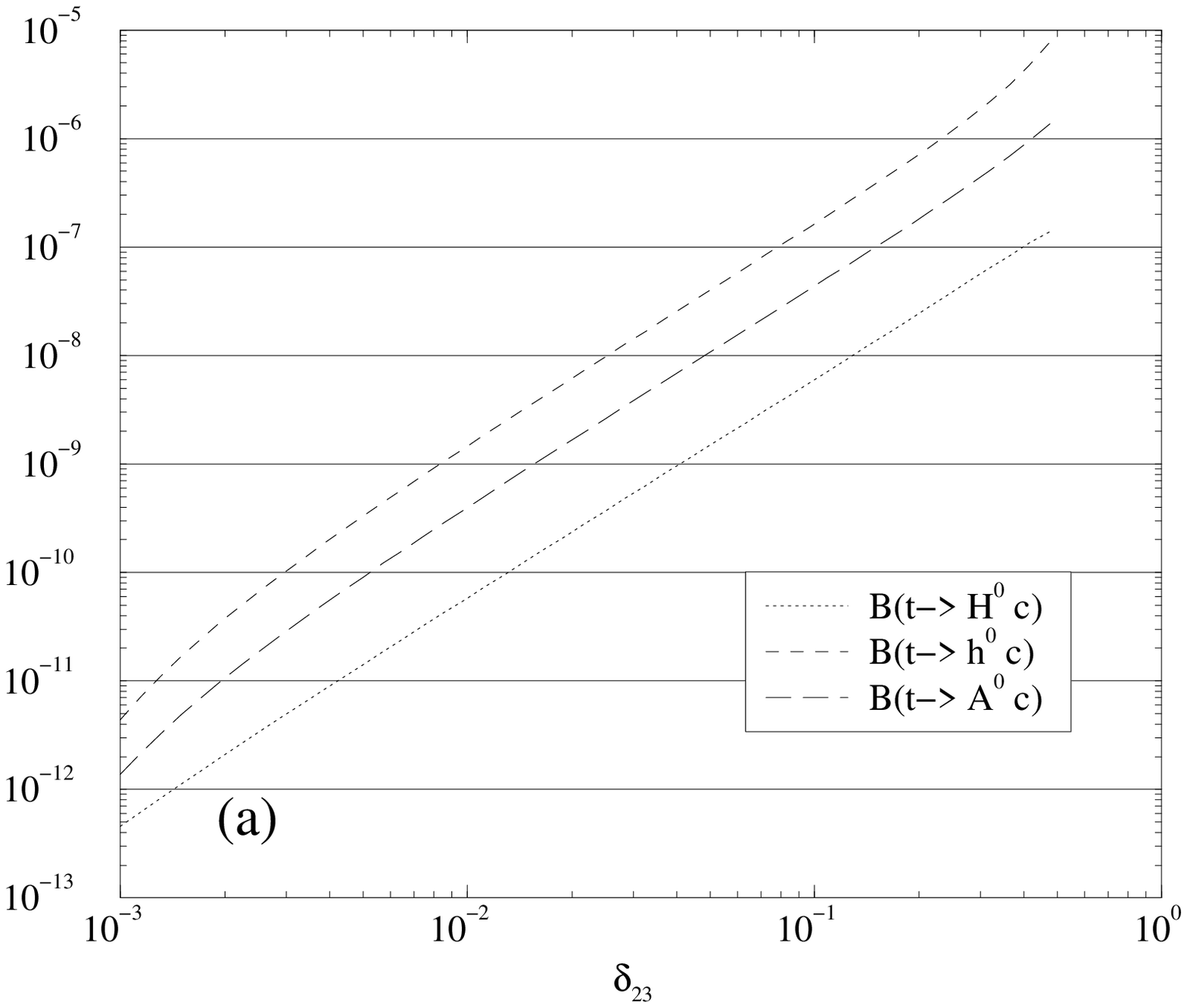}} &
\resizebox{5cm}{!}{\includegraphics{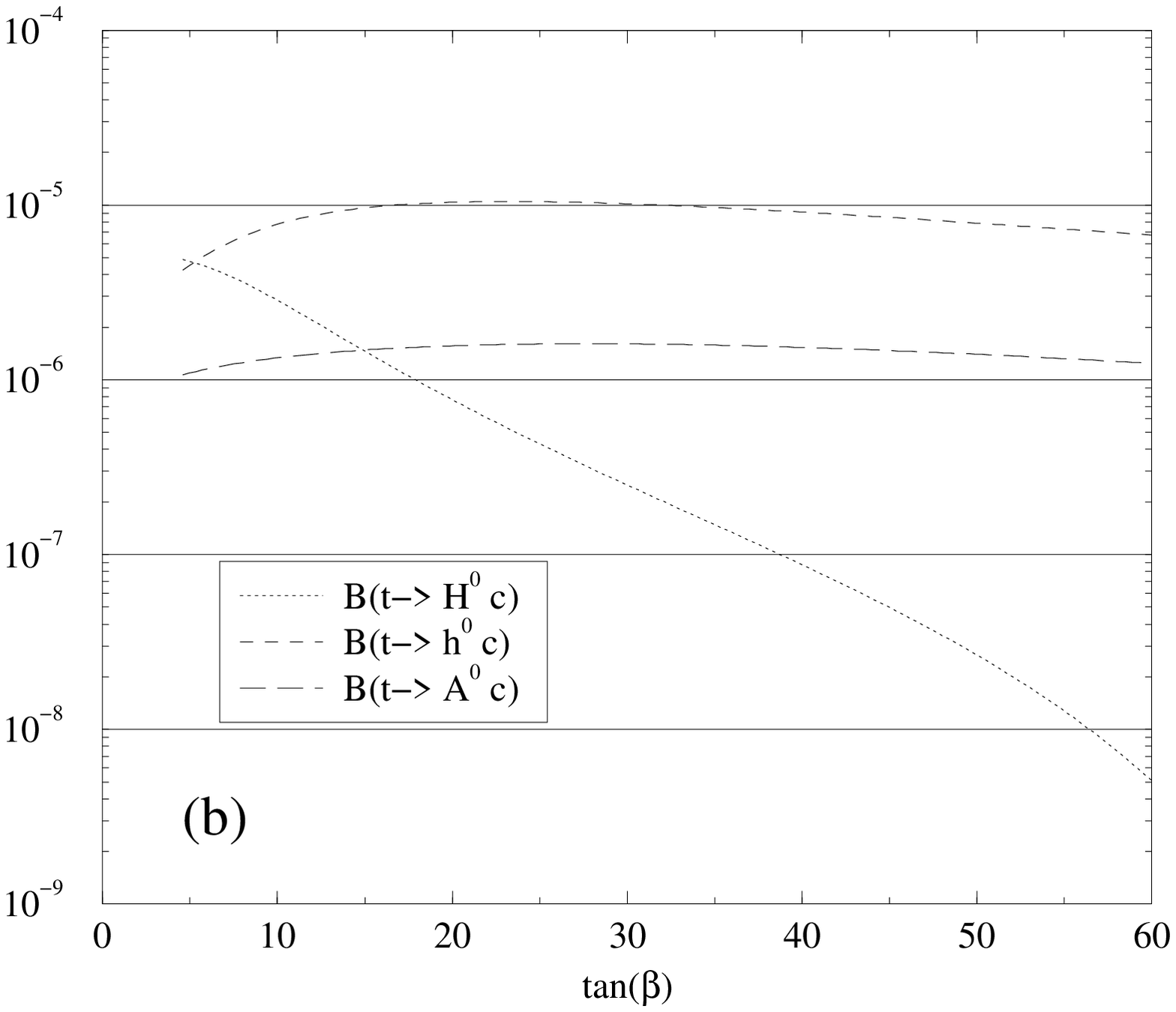}} \\
\resizebox{5cm}{!}{\includegraphics{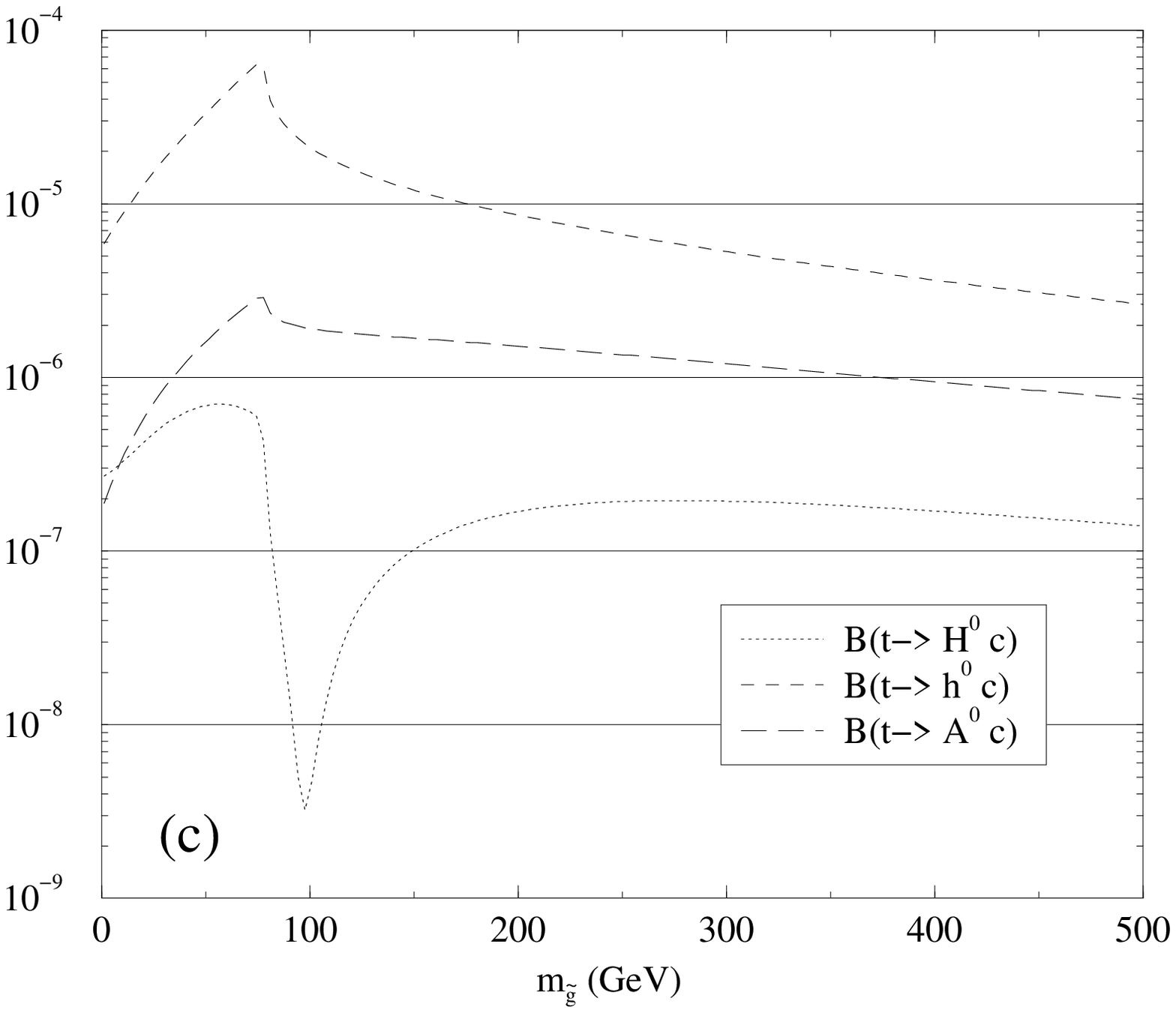}} &
\resizebox{5cm}{!}{\includegraphics{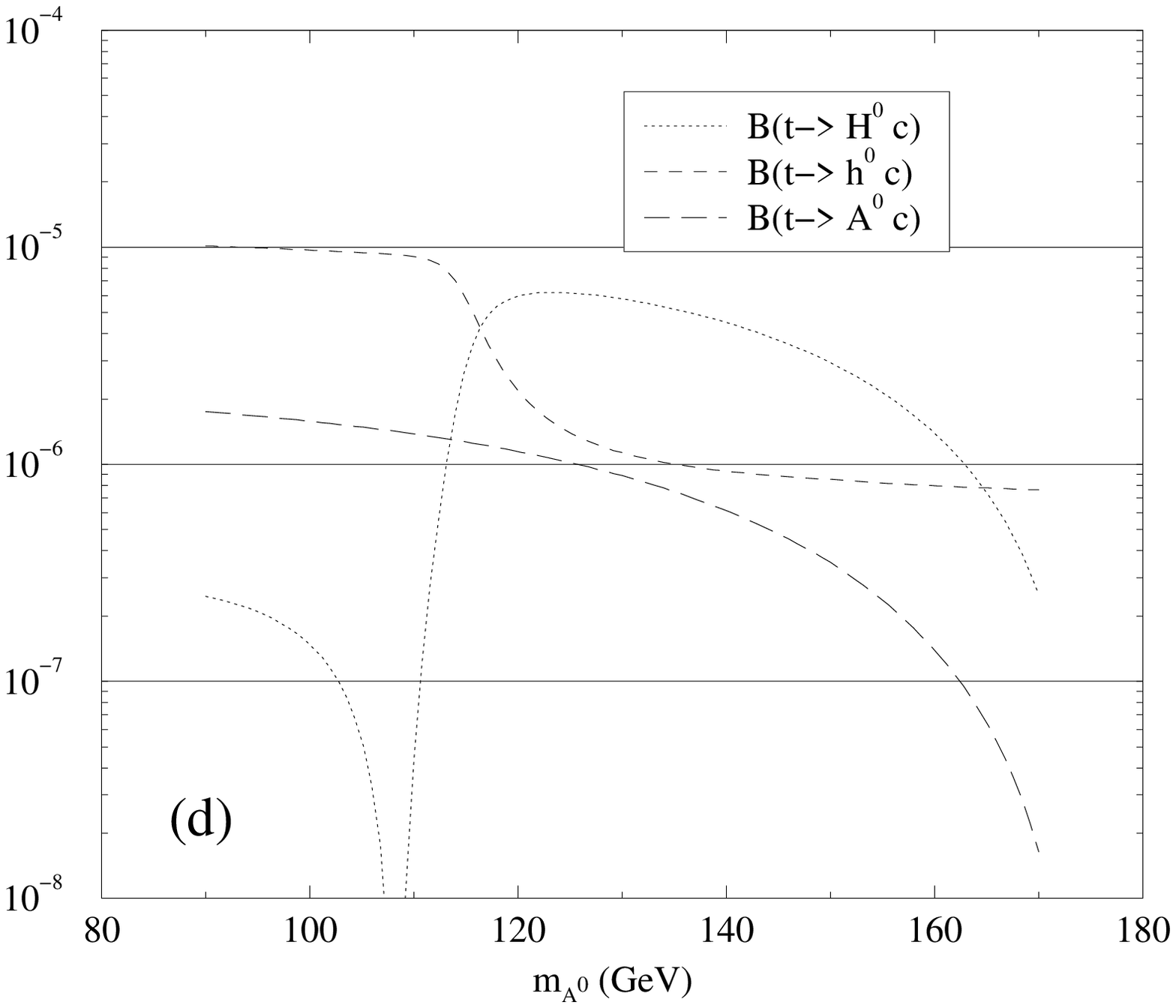}} 
\end{tabular}
\end{center}
\caption{Evolution of the SUSY-QCD contributions to the ratio (\ref{fiducialH}) with \textbf{(a)} the mixing parameter $\delta _{23}$ between
the 2nd and 3rd squark generations; \textbf{(b)} \tb; \textbf{(c)} the gluino mass $m_{\tilde{g}}$; and \textbf{(d)} the
pseudoscalar Higgs mass $M_{A^{0}}$. The rest of inputs are given in
eq.~({\ref{eq:inputqcd}})}\label{fig:mssmtch}
\end{figure}

Under the alignment hypothesis FCNC's are generated at the
one-loop level through the charged interactions of quarks with Higgs
bosons and squarks with charginos, that is, they are of electroweak (EW)
nature. In this case the largest rates are
driven by the trilinear scalar coupling $\tilde{d}_{a}\tilde{d}_{b}h$,
although the down-type-quark loops contributions are non-negligible. As
a consequence the largest FCNC decay rates are obtained at large \tb. 

Giving up alignment the leading FCNC rates are driven by means of the
SUSY-QCD tree-level vertex $u_a\tilde{u}_b\tilde{g}$ for $a\neq b$. The mixing terms
between generations are encoded in the parameter
 \begin{equation}
\delta_{ij}\equiv \frac{(M_{LL}^{2})_{ij}}{m_{i}\,m_{j}}\ \ (i\neq
j)\,\,,  \label{eq:defdelta}
\end{equation}
where $(M_{LL}^{2})_{ij}$ is the non-diagonal squared-mass-matrix
element between the $i$th and $j$th generations, and $m_{i}$ is the
mass parameter of the $i$th generation. Low energy experiments are used
to give upper bounds to the $\delta_{ij}$, but, whereas the mixing
between the 1st and 2nd generation are strongly restricted, the one
between the 2nd and the 3rd turns out to be basically
free~\cite{Gabbiani:1996hi}, 
and this is the one which has
a greatest impact in the process under study. We assume that
inter-generational mixing only exists between the left-handed squarks,
since this is the most natural
scenario~\cite{Duncan:1983iq}. 
A detailed analysis showed
that the presence of mixing in the right-handed squark sector does not
lead to a significant increase of the computed branching
ratios~\cite{deDivitiis:1997sh,Guasch:1999jp}.

Since the EW contributions lie about two orders of magnitude below
the SUSY-QCD ones~\cite{Guasch:1999jp}, we will concentrate in
the latter ones. In Fig.~\ref{fig:mssmtch} we present the fiducial
ratio~(\ref{fiducialH}) as a function of the most important parameters:
the mixing parameter between the 2nd and the 3rd generation
$\delta_{23}$, eq.~(\ref{eq:defdelta}); the gluino mass \mg; \tb; and
the mass of the pseudoscalar Higgs $m_{A^0}$. The set of reference
parameters used is:
\begin{equation}
\begin{array}{c}
\tb=35\,,\,\mA=100GeV\,,\,\mu=-200 GeV\,,\,A_t=A_q=-A_b=300
GeV\,,\,\\m_{\tilde t_1}=150 GeV\,,\,m_{\tilde
  q}\simeq200GeV\,,\,m_{\tilde g}=180 GeV\,,\,\delta_{12}=\delta_{13}=0.03\,,\,\delta_{23}=0.5\,.\\
\label{eq:inputqcd}
m_{t}=175\,GeV\,,\;m_{b}=5\,GeV\,,\,\,\alpha_{s}(m_{t})=0.11\,,\;\,
V_{cb}=0.040\,, 
\end{array}
\end{equation}
and the remaining ones are as in \cite{PDB2000}.

As anticipated, the most
important parameter is $\delta_{23}$. In Fig.~\ref{fig:mssmtch}a we see
that an increase in three orders of magnitude on $\delta_{23}$
corresponds to a change in six orders of magnitude on $B(t\to ch)$, a
fact that can be traced down to the quadratic dependence of the latter
on $\delta_{23}$. The dependence on \tb\  is rather mild since it
enters the amplitude through the $\tilde u_\alpha \tilde u_\beta h$
coupling as $1/\tb$, and also indirectly through the determination of
the squark masses. For the chosen set of parameters~(\ref{eq:inputqcd})
it has a non-negligible impact on the $H^0$ channel
(Fig.~\ref{fig:mssmtch}b). Although all Feynman diagrams proceed through
gluino exchange, the gluino mass turns out not to be a critical
quantity. In Fig.~\ref{fig:mssmtch}c we see that the decoupling of the
gluino is slow, a fact observed also in other Higgs bosons observables
related to the top quark~\cite{Coarasa:1996qa}. This fact can
be traced back to the presence of chirality-changing couplings, which
imply a corresponding gluino mass-insertion in the amplitude. Finally in
Fig.~\ref{fig:mssmtch}d we see that the smallest value for the $H^0$
branching ratio is not due to the smaller phase space available, but to
the value of the couplings. In fact $B(t\to cH^0)$ grows with  $\mA$
(and thus with $m_{H^0}$), until it dies out near the phase space 
kinematical limit.

\begin{figure}[t]
\begin{center}
\begin{tabular}{cc}
{\resizebox{5cm}{!}{\includegraphics{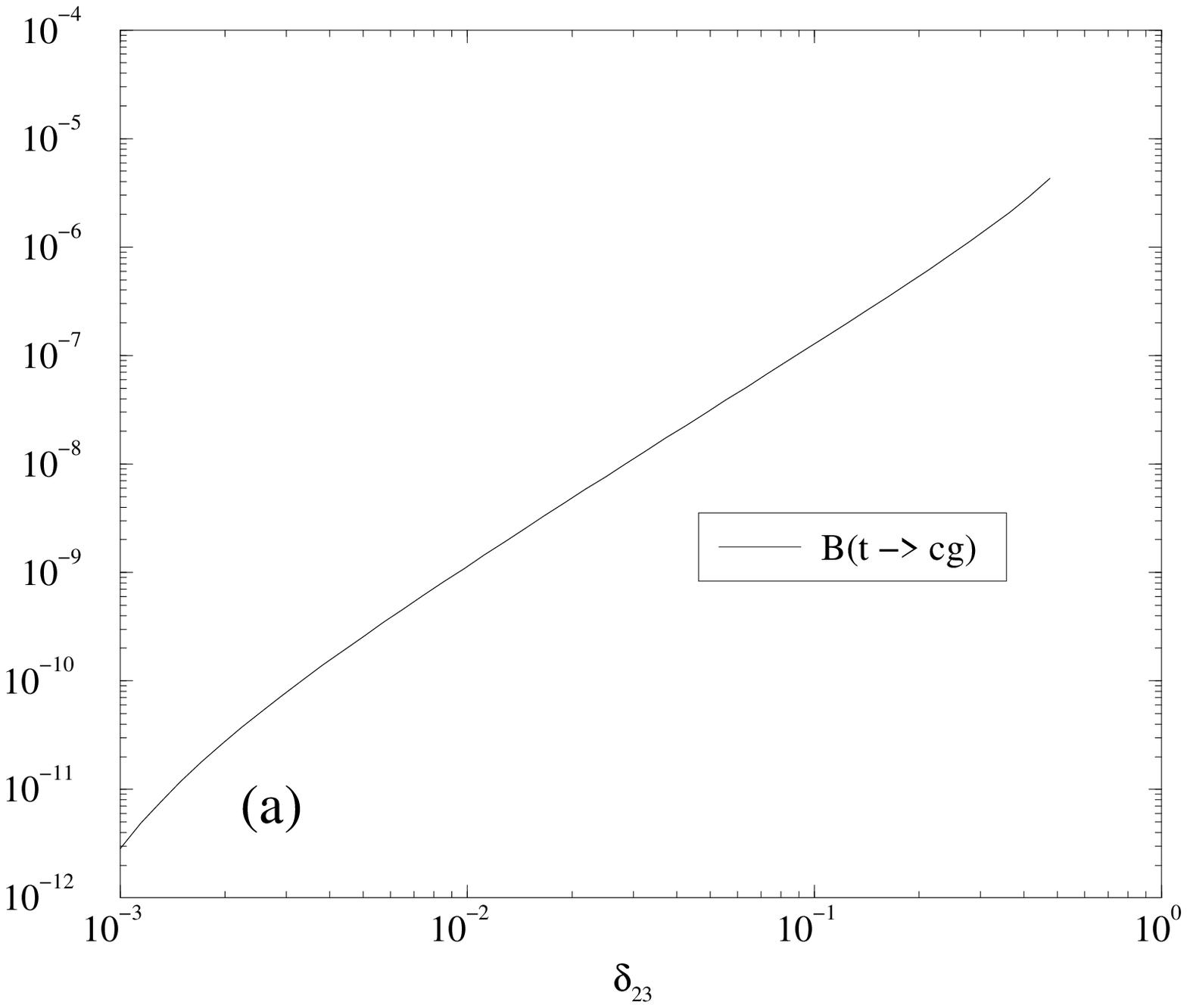}}}&
{\resizebox{5cm}{!}{\includegraphics{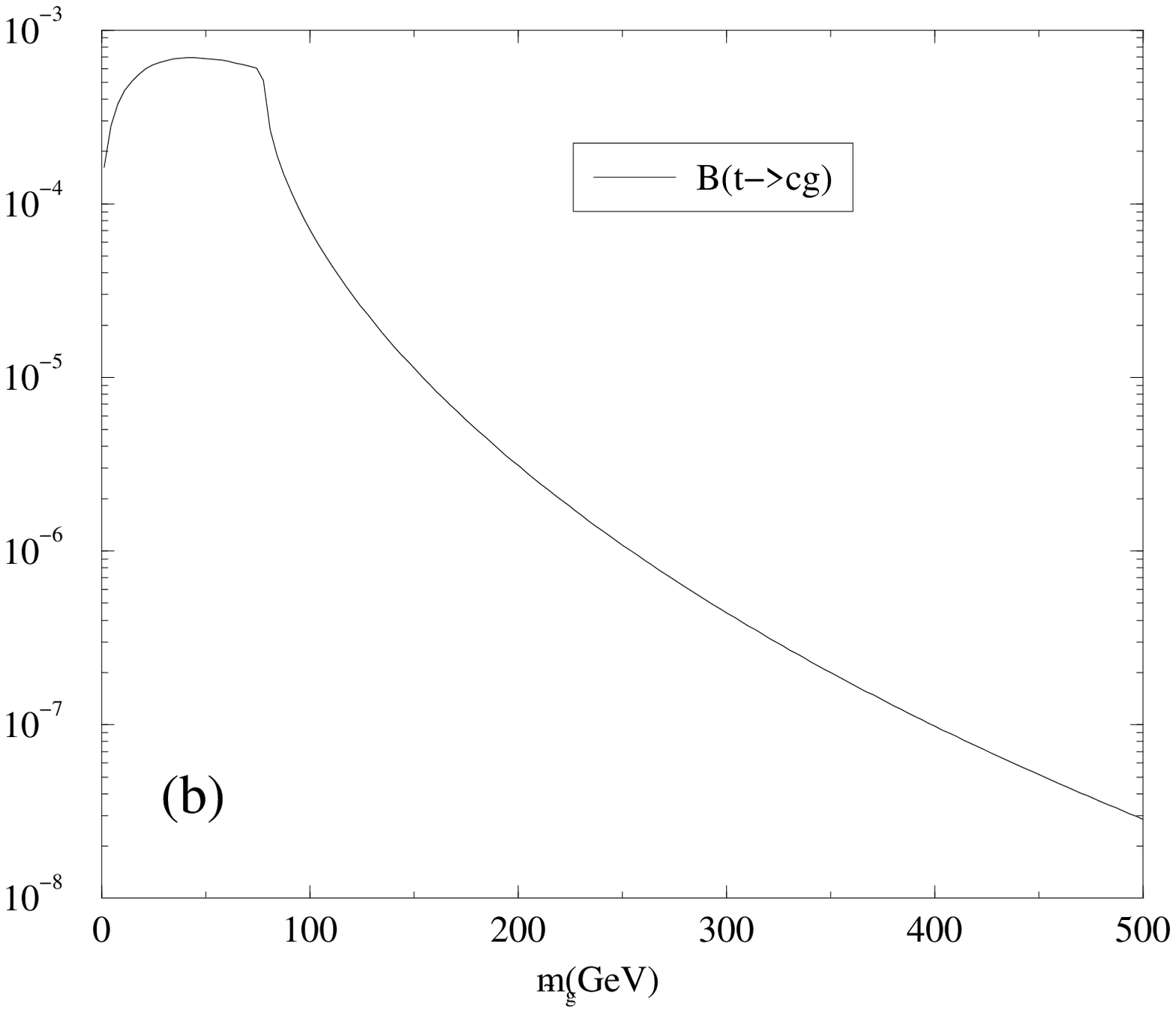}}}
\end{tabular}
\caption{Evolution of the SUSY-QCD effects on $B(t\rightarrow c\,g)$
as a function of \textbf{(a)} the mixing parameter $\delta _{23}$, and 
\textbf{(b)} the gluino mass $\mg$.}\label{fig:mssmtcg}
\end{center}
\end{figure}

In Fig.~\ref{fig:mssmtcg} we display the theoretical prediction for
$B(t\to cg)$ as a function of the gluino mass and $\delta_{23}$,
assuming $m_{H^\pm}>\mt$. The values for the ratio are below that of the
neutral Higgs bosons channels, but still some orders of magnitude above
the SM expected value for experimentally allowed values of 
$m_{\tilde  g}>180 GeV$. Again the branching ratio grows quadratically
with the mixing parameter $\delta_{23}$ (Fig.~\ref{fig:mssmtcg}a).   In
contrast with the Higgs bosons channels (Fig.~\ref{fig:mssmtch}c), the
gluon channel $B(t\to cg)$ shows a fast decoupling with 
the gluino mass (Fig.~\ref{fig:mssmtcg}b). 

\begin{figure}
\begin{center}
\begin{tabular}{cc}
{\resizebox{5cm}{!}{\includegraphics{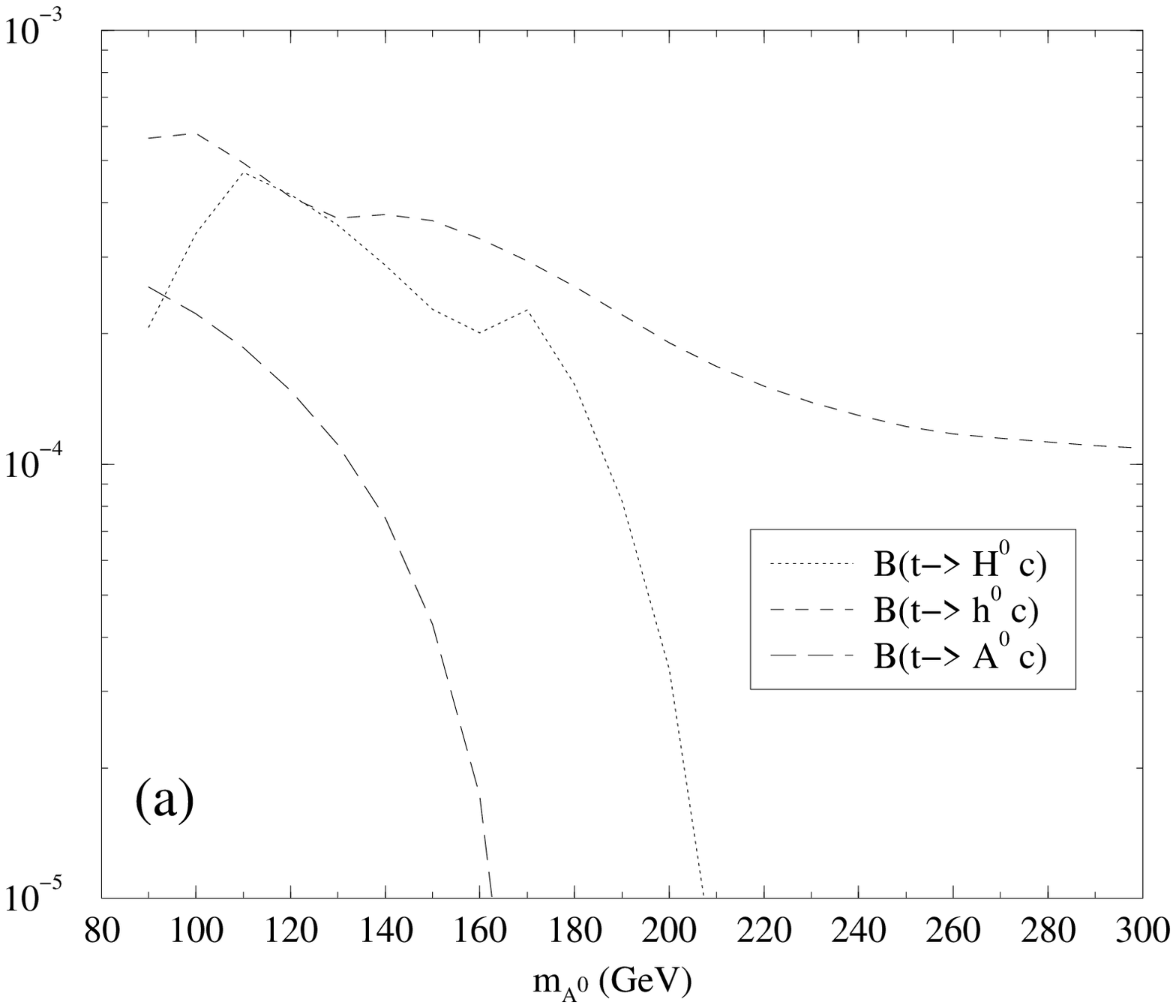}}}&
{\resizebox{5cm}{!}{\includegraphics{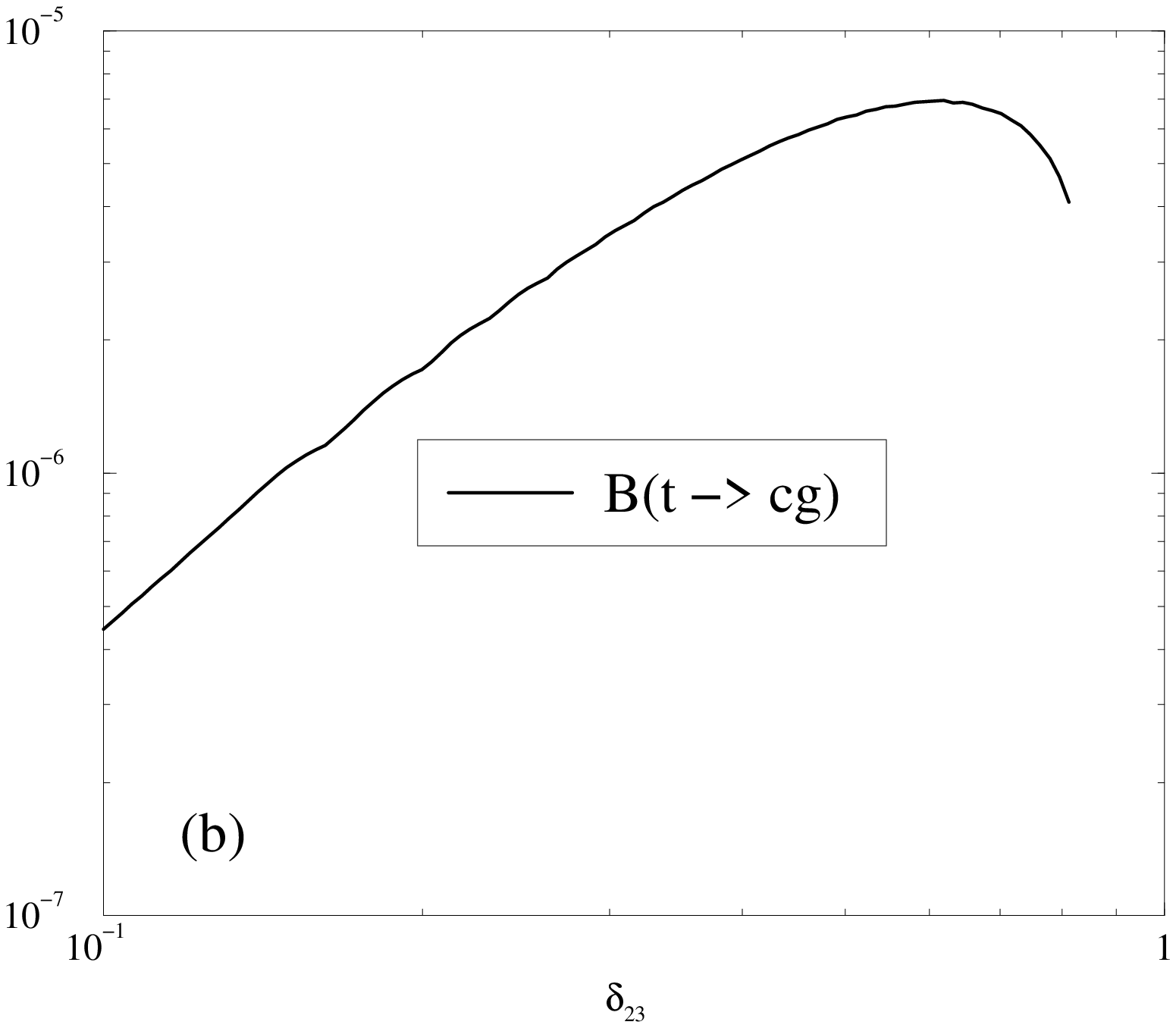}}}
\end{tabular}
\caption{Maximum value of $B(t\rightarrow c\,h)$ in the MSSM, obtained by
taking into account only the SUSY-QCD contributions, as a function of $\mA$; 
\textbf{(b)} maximum value of $B(t\rightarrow c\,g)$ as a
function of the intergenerational mixing parameter $\delta _{23}$ in the LH
sector. In all cases the scanning for the rest of parameters of the MSSM has
been performed within the phenomenologically allowed region.}\label{fig:maximmssm}
\end{center}
\end{figure}

In Fig.~\ref{fig:maximmssm} we present the maximum rates for 
$B(t\to ch)$ and $B(t\to cg)$ -- eq.~(\ref{fiducialH}) -- in the MSSM. We have made a comprehensive scan of
the MSSM parameter space, taking into account present constraints from
experiment. Perhaps the most noticeable result is that the decay into the
lightest  
MSSM Higgs boson ($t\rightarrow c\,h^{0}$) is the one that can be maximally
enhanced, and reaching values of order 
$B^{\mathrm{MSSM}}(t\rightarrow c\,h^{0})\sim 10^{-4}$ that stay fairly stable
all over the parameter space. The FCNC top quark
decay into the lightest Higgs scalar can have an observable ratio in a large
portion of the parameter space, and in particular for almost all the range
of Higgs boson masses. Needless to say, not all of the maxima can be
simultaneously attained as they are obtained for different values of the
parameters. The maximum FCNC rate of the gluon channel in the MSSM reads
(Fig.~\ref{fig:maximmssm}b)
$B^{\mathrm{MSSM}}({t\rightarrow c\,g})\lsim \,10^{-5}$,
but it never really reaches the critical value $10^{-5}$, which
  can be considered as the visible threshold for the next generation of 
colliders (see Sec.~\ref{sec:discussion}).
\section{$t\to c h$ and $t\to c g$ in the general 2HDM} 

In the 2HDM case a  highly relevant parameter is $\tan\beta$, which must 
be restricted to the approximate range
\begin{equation} 
0.1<\tan\beta\stackm60  \label{tbrange} 
\end{equation} 
in perturbation theory.
It is to be expected from the various couplings 
involved in the processes under consideration that the low $\tan\beta$ 
region could be relevant for both the Type~I and Type~II 2HDM's. In 
contrast, the high $\tan\beta$ region is only potentially important for the 
Type~II. However, the eventually relevant regions of parameter space are 
also determined by the value of the mixing angle $\alpha$, as we shall see 
below. 
 
Of course there are several restrictions that must be respected by our 
numerical analysis. 
First, the 
one-loop corrections to the $\rho$-parameter from the 2HDM sector cannot 
deviate from the reference SM contribution in more than one per mil~\cite{PDB2000}:  
$|\delta\rho^{2HDM}|\leqslant0.001$. From the analytical expression for
$\delta\rho$ in the general 2HDM we have  
introduced this numerical condition in our codes. 

For both models we have imposed the condition that the (absolute 
value) of the trilinear Higgs self-couplings do not exceed the maximum 
unitarity limit tolerated for the SM trilinear coupling:  
$\left| \lambda_{HHH}\right| \leqslant\left| 
\lambda_{HHH}^{(SM)}(m_{H}=1\,TeV)\right| ={3\,g\,(1\,TeV)^{2}}/({2\,M_{W}})$. 
In the MSSM case it was not necessary to impose this restriction because
the Higgs self-couplings are purely gauge. As for the $\delta\rho$
constraint in the MSSM, we have checked that it is satisfied. It
is in the 
2HDM case that one has to keep an eye very seriously on $\delta\rho$
because it grows with the Higgs boson mass squared differences, whereas
in the MSSM the mass differences are much more tamed in all sectors of
the theory.
\begin{figure}[t] 
\begin{center} 
\resizebox{14cm}{!}{\includegraphics{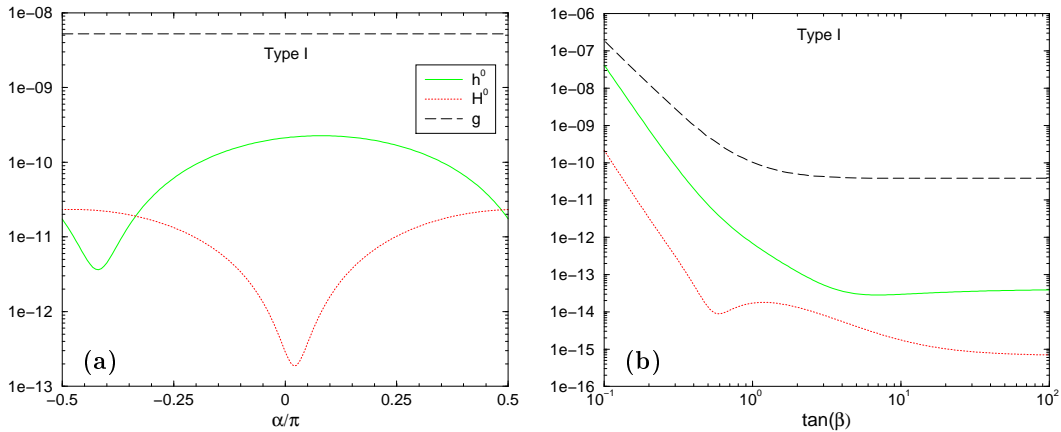}} 
\end{center} 
\caption{Evolution of the FCNC top quark fiducial
  ratio~(\ref{fiducialH}) -- and the corresponding one for $B(t\to g+c)$
  -- in Type~I 2HDM versus: \textbf{(a)} the mixing angle
  $\protect\alpha$ in the CP-even Higgs sector, in units of
  $\protect\pi$;   
\textbf{(b)} $\tan\protect\beta$. The values of the fixed parameters are as 
in eqs.~(\ref{inputsmixing}) and (\ref{inputsmasses}).} 
\label{fig:2} 
\end{figure} 
 
\begin{figure}[t]
\begin{center} 
\resizebox{14cm}{!}{\includegraphics{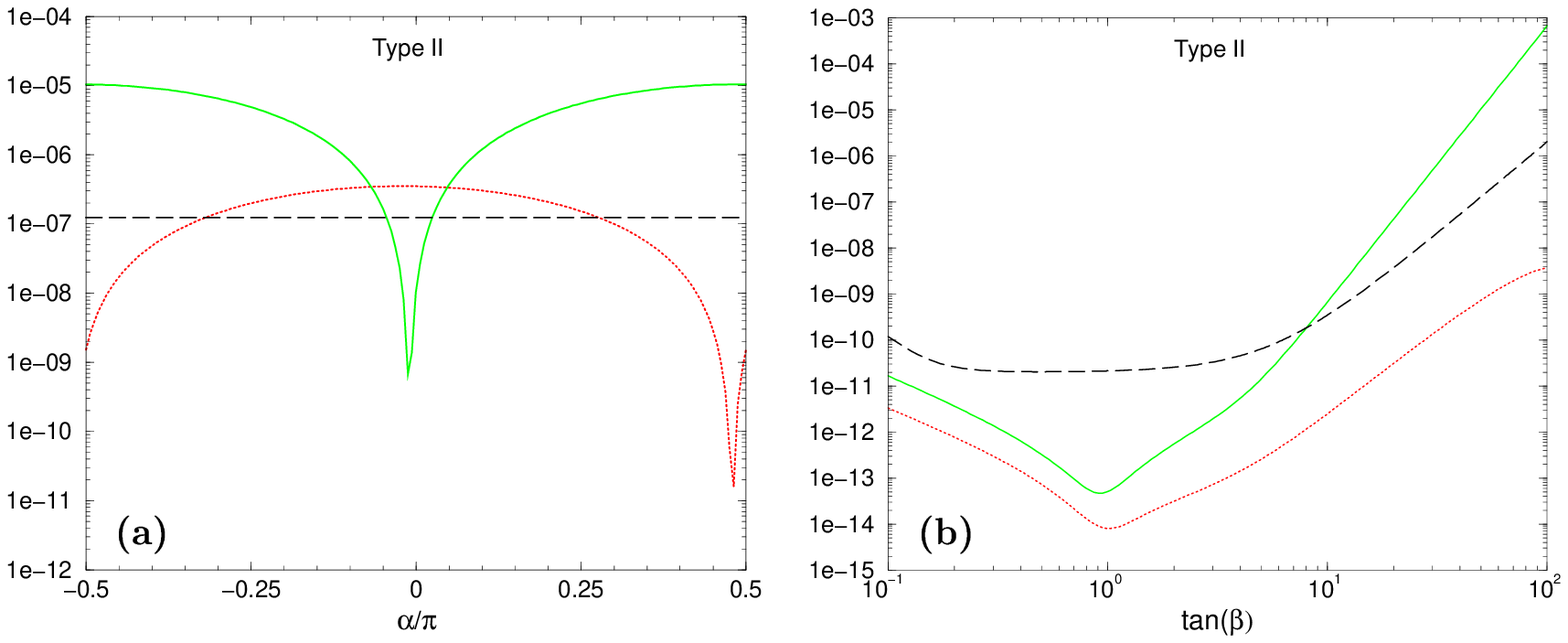}} 
\end{center} 
\caption{As in Fig.~\ref{fig:2}, but for the 2HDM~II. The plot in (b) 
continues above the bound in eq.~(\ref{tbrange}) just to better show the 
general trend.} 
\label{fig:3} 
\end{figure}

\begin{figure}[t] 
\begin{center} 
\resizebox{14cm}{!}{\includegraphics{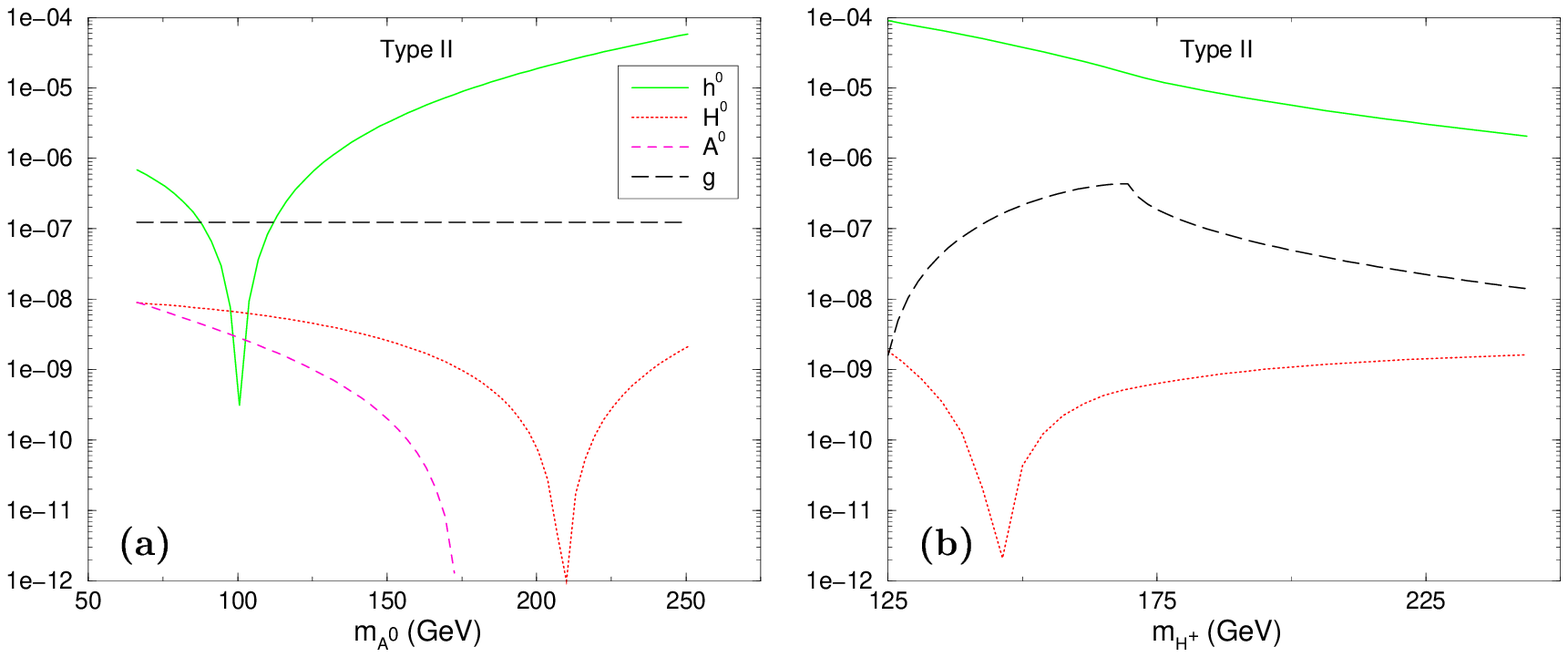}} 
\end{center}
\caption{Evolution of the FCNC top quark fiducial
  ratios~(\ref{fiducialH}) -- and the corresponding one for $B(t\to
  g+c)$ -- in Type~II 2HDM versus: \textbf{(a)} the CP-odd Higgs boson 
mass $m_{A^{0}}$; \textbf{(b) }the charged Higgs boson mass $m_{H^{\pm}}$. 
The values of the fixed parameters are as in eqs.~(\ref{inputsmixing})
  and (\ref{inputsmasses}). The plot in (b) starts below the bound
  $m_{H^{\pm}}>165\,GeV$ mentioned in the text to better show the
  general  
trend. } 
\label{fig:4} 
\end{figure}

\begin{figure}[t] 
\begin{center} 
\resizebox{14cm}{!}{\includegraphics{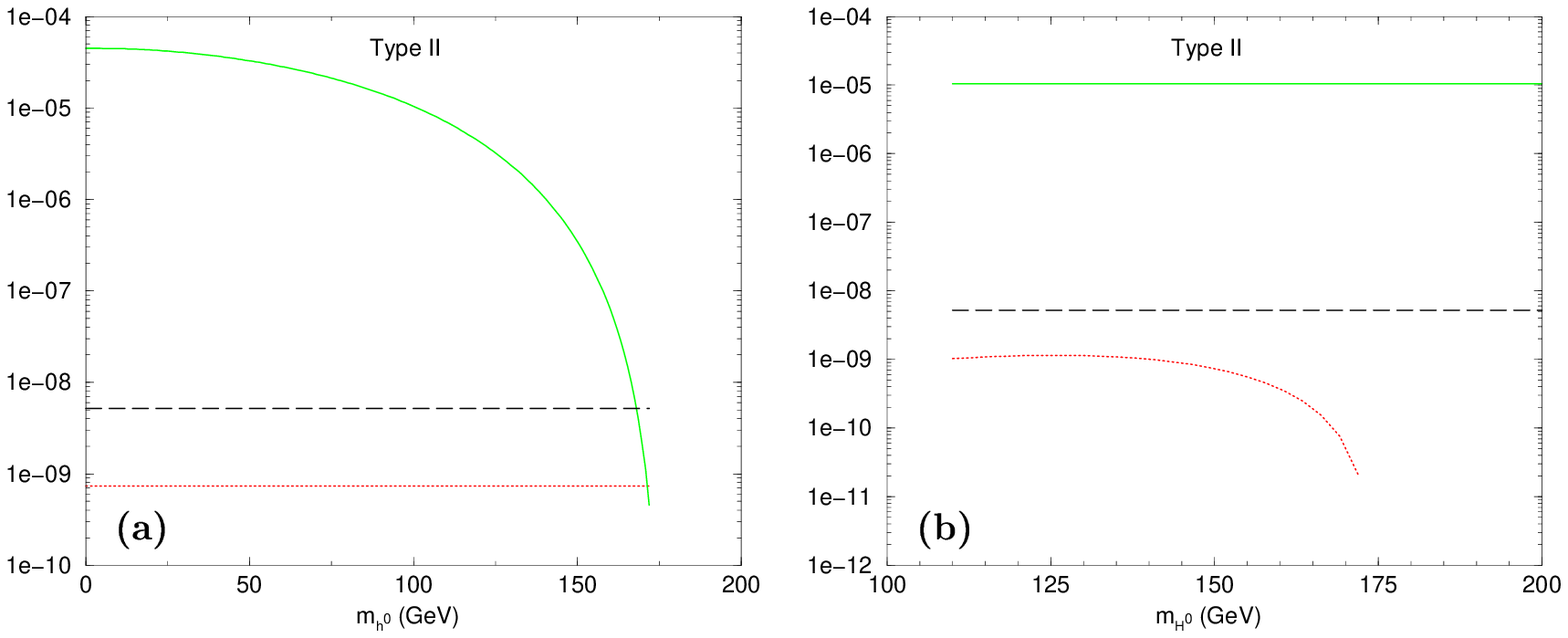}} 
\end{center} 
\caption{As in Fig.~\ref{fig:4}, but plotting versus: \textbf{(a) }the 
lightest CP-even Higgs boson mass $m_{h^{0}}$; \textbf{(b) }the heaviest 
CP-even Higgs boson mass $m_{H^{0}}$.} 
\label{fig:5} 
\end{figure}

The combined set of independent conditions turns out to be quite effective 
in narrowing down the permitted region in the parameter space, as can be 
seen in Figs.~\ref{fig:2}-\ref{fig:5} where we plot the fiducial FCNC
rate~(\ref{fiducialH}) and the corresponding one for the gluon channel
versus the parameters~(\ref{freeparam}).  
The cuts in some of these curves just reflect the fact that at least one of 
these conditions is not fulfilled. 
 
After scanning the parameter space, we see in Figs.~\ref{fig:2}-\ref{fig:3} 
that the 2HDM~I (resp. 2HDM~II) prefers low values (resp. high values)
of $\tan\alpha$ and $\tan \beta$ for a given channel, e.g. $t\rightarrow  
h^{0}\,c $. Therefore, the following choice of mixing angles will be made to 
optimize the presentation of our numerical results:  
\begin{equation} 
\text{2HDM I} :\ \tan\alpha=\tan\beta=1/4\,\,\,;  \,\,\,
\text{2HDM II} :\ \tan\alpha=\tan\beta=50\,.  \label{inputsmixing} 
\end{equation}

We point out that, for the same values of the masses, one obtains the same 
maximal FCNC rates for the alternative channel $t\rightarrow H^{0}\,c$ 
provided one just substitutes
$\alpha\rightarrow\pi/2-\alpha$. Equation~(\ref{inputsmixing}) defines
the eventually relevant regions of parameter  
space and, as mentioned above, depend on the values of the mixing angles
$\alpha$ and $\beta$, namely $\beta \simeq\alpha\simeq0$ for Type~I and
$\beta\simeq\alpha\simeq\pi/2$ for Type~II.  
Despite naive expectations, and due to the structure of the Yukawa
couplings of Type II models, there is a cancellation of the
contributions in the low $\tan\beta$ end. This is in contrast to Type I
models where the maximal ratios occur. So the most favoured region of
Type II models definitely is the high $\tan\beta$ one.

Due to the $\alpha \rightarrow \pi /2-\alpha $ symmetry of the maximal rates 
for the CP-even Higgs channels, it is enough to concentrate the numerical 
analysis on one of them, but one has to keep in mind that the other channel 
yields the same rate in another region of parameter space. Whenever a mass 
has to be fixed, we choose conservatively the following values for both 
models:
\begin{equation} 
m_{h^{0}}=100\,GeV\,,\;m_{H^{0}}=150\,GeV\,,\;\;m_{A^{0}}=m_{H^{\pm 
}}=180\,GeV\,.  \label{inputsmasses} 
\end{equation} 
The variation 
of the results with respect to the masses is studied in Figs.~\ref{fig:4}-\ref{fig:5}. In particular, in Fig.~\ref{fig:4} we can see the (scanty) rate 
of the channel $t\rightarrow A^{0}\,c$ when it is kinematically
allowed. 
This is easily understood as it is the only one that does not have trilinear 
couplings with the other Higgs particles. While it 
does have trilinear couplings involving Goldstone bosons, these are not 
enhanced. The crucial role played by the trilinear Higgs self-couplings in 
our analysis cannot be underestimated as they can be enhanced by playing 
around with both (large or small) $\tan \beta $ \emph{and} also with the 
mass splittings among Higgses. This feature is particularly clear in Fig.~\ref{fig:4}a where the rate of the channel $t\rightarrow h^{0}\,c$ is 
dramatically increased at large $m_{A^{0}}$, for fixed values of the other 
parameters and preserving our list of constraints. 

From Figs.~\ref{fig:2}a and \ref{fig:2}b it is pretty clear that the 
possibility to see FCNC decays of the top quark into Type~I Higgs bosons is 
plainly hopeless even in the most favorable regions of parameter space -- 
the lowest (allowed) $\tan\beta$ end. In fact, the highest rates remain 
neatly down $10^{-6}$, and therefore they are (at least) one order of 
magnitude below the threshold sensibility of the best high luminosity top 
quark factory in the foreseeable future (see Section~\ref{sec:discussion}). 

Fortunately, the meager situation just described does not replicate for Type 
II Higgs bosons. For, as shown in Figs.~\ref{fig:3}a and \ref{fig:3}b, the 
highest potential rates are of order $10^{-4}$, and so there is hope for 
being visible. In this case the most favorable region of parameter space is 
the high $\tan\beta$ end in eq.~(\ref{tbrange}). Remarkably, there is no 
need of risking values over and around $100$ 
to obtain the desired rates. But 
it certainly requires to resort to models whose hallmark is a large value of  
$\tan\beta$ of order or above $m_{t}/m_{b}\stackM35$. As for the dependence 
of the FCNC rates on the various Higgs boson masses
(Cf. Figs.~\ref{fig:4}-\ref{fig:5}) we see that for large $m_{A^{0}}$
the decay $t\rightarrow  
h^{0}\,c$ can be greatly enhanced as compared to $t\rightarrow g\,c$.
We also note (from the 
combined use of Figs.~\ref{fig:3}b, \ref{fig:4}a and \ref{fig:4}b) that in 
the narrow range where $t\rightarrow H^{+}\,b$ could still be open in the 
2HDM~II, the rate of $t\rightarrow h^{0}\,c$ becomes the more visible the 
larger and larger is $\tan\beta$ and $m_{A^{0}}$. Indeed, in this region one 
may even overshoot the $10^{-4}$ level without exceeding the upper
bound~(\ref{tbrange}) while also keeping under control the remaining
constraints. 
Finally, the evolution of the rate~(\ref{fiducialH}) and $B(t\to g+c)$
with respect to the two CP-even Higgs boson  
masses is shown in Figs.~\ref{fig:5}a and \ref{fig:5}b.

\section{Discussion and conclusions} 
\label{sec:discussion}
 In the near and middle future, with the upgrades of the Tevatron (Run II, 
TeV33), the advent of the LHC, and the construction of an $e^{+}e^{-}$ 
linear collider (LC), new results on top quark physics, 
and possibly also on 
Higgs physics, will be obtained.
With datasets from LHC and LC 
increasing to several~$100 fb^{-1}/$year in the high-luminosity phase, one 
should be able to pile up an enormous wealth of statistics on top quark 
decays. Therefore, these machines should be very useful to 
analyze rare decays of the top quark, viz. decays whose branching fractions 
are extremely small ($\stackm10^{-5}$). 

The sensitivities to FCNC top quark decays for $100\,fb^{-1}$ of integrated 
luminosity in the relevant colliders are estimated to be~\cite{Frey:1997sg}:  
\begin{equation} \begin{array}{l}
\mathrm{\mathbf{LHC:}}B(t\rightarrow c\,X)  \gsim5\times10^{-5}\,\,,\,\, 
\mathrm{\mathbf{LC:}}B(t\rightarrow c\,X)  \gsim5\times10^{-4}\,, \\
\mathrm{\mathbf{TEV33:}}B(t\rightarrow c\,X)  \gsim5\times10^{-3}\,\,\,.  
\end{array}\label{sensitiv}
\end{equation} 
This estimation has been confirmed by a full signal-background analysis
for the hadron colliders 
and also
for the LC in the case of gauge boson
decays~\cite{Aguilar-Saavedra:2000aj}. 
From these experimental expectations and our numerical results it becomes 
patent that whilst the Tevatron will remain essentially blind to this kind 
of physics, the LHC and the LC will have a significant potential to observe 
FCNC decays of the top quark beyond the SM. Above all there is a possibility 
to pin down top quark decays into neutral Higgs particles,
eq.~(\ref{Higgschannels}), within the framework of the general 2HDM~II
provided $\tan\beta\stackM\,m_{t}/m_{b}\sim35$, and within the MSSM
provided~$\delta_{23}$--eq.~(\ref{eq:defdelta})-- is large. The maximum
rates are of 
order $10^{-4}$ in both models and correspond to the two CP-even
scalars. In the MSSM the lightest Higgs boson is highlighted all over
the $\mA$ range. This
conclusion is  
remarkable from the practical (quantitative) point of view, and also 
qualitatively because the top quark decay into the SM Higgs particle is
the less favorable top 
quark FCNC rate in the SM. On the other hand, we deem practically hopeless 
to see FCNC decays of the top quark in a general 2HDM~I for which the 
maximum rates are of order $10^{-7}$. This order of magnitude cannot be 
enhanced unless one allows $\tan\beta\ll0.1$, but the latter possibility is 
unrealistic because perturbation theory breaks down and therefore one cannot 
make any prediction within our approach. 
 
We have made a parallel numerical analysis of the gluon channel $t\rightarrow c\;g$. We confirm that this is another 
potentially important FCNC mode of the top quark in extensions of the 
SM~\cite{Eilam:1991zc,deDivitiis:1997sh,Guasch:1999jp,Bejar:2000ub} but,
unfortunately, it still falls a bit too short to be  
detectable. The maximum rates for this channel lie below $10^{-6}$ in the 
2HDM~I (for $\tan\beta>0.1)$ and in the 2HDM~II (for $\tan\beta<60$),
and below $10^{-5}$ for the MSSM, and so 
it will be hard to deal with it even at the LHC. 
 
We are thus led to the conclusion that the Higgs channels~(\ref{Higgschannels}), 
more specifically the CP-even ones, give the highest potential rates for top 
quark FCNC decays in a general 2HDM~II and the MSSM. Most significant of all: they are  
\emph{the only} FCNC decay modes of the top quark, within the simplest 
renormalizable extensions of the SM, that have a real chance to be seen in 
the next generation of high energy, high luminosity, colliders.

Although the 2HDM~II and the MSSM show similar behaviour, there exist
some conspicuous differences on which we wish 
to elaborate a bit in what follows. First, in the general 
2HDM~II the two channels $t\rightarrow(h^{0},H^{0})\;c$ give the same 
maximum rates, provided we look at different (disjoint) regions of the 
parameter space. The $t\rightarrow A^{0}\;c$ channel is, as mentioned, 
negligible with respect to the CP-even modes. Hereafter we will discard this 
FCNC top quark decay mode from our discussions within the 2HDM context. On 
the other hand, in the MSSM there is a most distinguished channel, 
viz.~$t\rightarrow h^{0}\;c$, which can be high-powered by the SUSY stuff all over 
the parameter space. In this framework the mixing angle $\alpha$ becomes 
stuck once $\tan\beta$ and the rest of the independent parameters are given, 
and so there is no possibility to reconvert the couplings between $h^{0}$ 
and $H^{0}$ as in the 2HDM. Still, we must emphasize that in the MSSM the 
other two decays $t\rightarrow H^{0}\;c$ and $t\rightarrow A^{0}\;c$ can be 
competitive with $t\rightarrow h^{0}\;c$ in certain portions of parameter 
space. For example, $t\rightarrow H^{0}\;c$ becomes competitive when the 
pseudoscalar mass is in the range $110\,GeV<m_{A^{0}}\stackm170\,GeV$
--Cf. Fig.~\ref{fig:mssmtch}d. The possibility of having more than one FCNC decay~(\ref{Higgschannels}) near the visible level is a feature which is virtually impossible in the 
2HDM~II. Second, the reason why $t\rightarrow h^{0}\;c$ in the MSSM is so 
especial is that it is the only FCNC top quark decay~(\ref{Higgschannels}) 
which is always kinematically open throughout the whole MSSM parameter 
space, while in the 2HDM all of the decays~(\ref{Higgschannels}) could be, 
in the worst possible situation, dead closed. Nevertheless, this is not the 
most likely situation in view of the fact that all hints from high precision 
electroweak data seem to recommend the existence of (at least) one 
relatively light Higgs boson~\cite{TJunk,Hagiwara}. This is certainly an 
additional motivation for our work, as it leads us to believe that in all 
possible (renormalizable) frameworks beyond the SM, and not only in SUSY, we 
should expect that at least one FCNC decay channel~(\ref{Higgschannels}) 
could be accessible. Third, the main origin of the maximum FCNC rates in the 
MSSM traces back to the tree-level FCNC couplings of the gluino~\cite{Guasch:1999jp}. 
These are strong couplings, and moreover they are very weakly restrained by 
experiment. In the absence of such gluino couplings, or perhaps by further 
experimental constraining of them in the future, the FCNC rates in the MSSM 
would boil down to just the EW contributions, to wit, those 
induced by charginos, squarks and also from SUSY Higgses. The associated 
SUSY-EW rate is of order~$10^{-6}$ at most~\cite{Guasch:1999jp}, and
therefore it is barely  
visible, most likely hopeless even for the LHC. In contrast, in the general 
2HDM the origin of the contributions is purely EW and the maximum rates are 
two orders of magnitude higher than the full SUSY-EW effects in the MSSM. It 
means that we could find ourselves in the following situation. Suppose that 
the FCNC couplings of the gluino get severely restrained in the future and 
that we come to observe a few FCNC decays of the top quark into 
Higgs bosons, perhaps at the LHC and/or the LC. Then we would immediately 
conclude that these Higgs bosons could not be SUSY-MSSM, whilst 
they could perhaps be CP-even members of a 2HDM~II. Fourth, the gluino 
effects are basically insensitive to $\tan\beta$, implying that the maximum 
MSSM rates are achieved equally well for low, intermediate or high values of  
$\tan\beta,$ whereas the maximum 2HDM~II rates (comparable to the MSSM ones) 
are attained only for high $\tan\beta$.

The last point brings about the  question of  whether it
would be possible to discern between different models if these decays
are detected. The answer is, most  likely yes.
There are many possibilities and corresponding strategies, but we will limit 
ourselves to point out some of them. For example, let us consider the type 
of signatures involved in the tagging of the Higgs channels. In the favorite 
FCNC region~(\ref{inputsmixing}) of the 2HDM~II, the combined decay  $t\rightarrow 
h\;\,c\rightarrow cb\overline{b}$ is possible only for $h^{0}$ or for
$H^{0}$, but not for both -- Cf. Fig.~\ref{fig:3}a -- whereas in the
MSSM, $h^{0}$  
together with $H^{0}$, are highlighted for $110\,GeV<m_{A^{0}}<m_{t}$, with 
no preferred $\tan \beta $ value. And similarly, $t\rightarrow A^{0}\;c$ is 
also non-negligible for $m_{A^{0}}\stackm120\,GeV$ --Cf. Fig.~\ref{fig:mssmtch}d. Then 
the process $t\rightarrow h\;\,c\rightarrow cb\overline{b}$ gives 
rise to high $p_{T}$ charm-quark jets and a recoiling $b\overline{b}$ pair 
with large invariant mass. It follows that if more than one distinctive 
signature of this kind would be observed, the origin of the hypothetical 
Higgs particles could not probably be traced back to a 2HDM~II.  
 
One might worry that in the case of $h^{0}$ and $H^{0}$ they could also (in 
principle) decay into electroweak gauge boson pairs $h^{0},H^{0}\rightarrow 
V_{ew}\overline{V}_{ew}$, which in some cases could be kinematically 
possible. But this is not so in practice for the
2HDM~II~\cite{Bejar:2000ub}.
Again, at variance with this situation, 
in the MSSM case $H^{0}\rightarrow V_{ew}\overline{V}_{ew}$ is perfectly 
possible -- not so $h^{0}\rightarrow V_{ew}\overline{V}_{ew}$ due to the 
aforementioned upper bound on $m_{h^{0}}$ -- because $\tan \beta $ has no 
preferred value in the most favorable MSSM decay region of $t\rightarrow 
H^{0}\;\,c$. Therefore, detection of a high $p_{T}$ charm-quark jet against 
a $V_{ew}\overline{V}_{ew}$ pair of large invariant mass could only be 
advantageous in the MSSM, not in the 2HDM. Similarly, for $\tan \beta \stackM%
1$ the decay $H^{0}\rightarrow h^{0}\;h^{0}$ (with real or virtual $h^{0}$) 
is competitive in the MSSM
in a region where the parent FCNC 
top quark decay is also sizeable. Again this is impossible in the 2HDM~II 
and therefore it can be used to distinguish the two (SUSY and non-SUSY) 
Higgs frames. 
 
Finally, even if we place ourselves in the high $\tan\beta$ region both for 
the MSSM and the 2HDM~II, then the two frameworks could still possibly be 
separated provided that two Higgs masses were known, perhaps one or both of 
them being determined from the tagged Higgs decays themselves, eq.~(\ref 
{Higgschannels}). Suppose that $\tan\beta$ is numerically known (from other 
processes or from some favorable fit to precision data), then the full 
spectrum of MSSM Higgs bosons would be approximately determined (at the tree 
level) by only knowing one Higgs mass, a fact that could be used to check 
whether the other measured Higgs mass becomes correctly predicted. Of 
course, the radiative corrections to the MSSM Higgs mass relations can be 
important at high $\tan\beta$~\cite{Carena:1996wu},
but these could be taken into 
account from the approximate knowledge of the relevant sparticle masses 
obtained from the best fits available to the precision measurements within 
the MSSM. If there were significant departures between the predicted mass 
for the other Higgs and the measured one, we would probably suspect that the 
tagged FCNC decays into Higgs bosons should correspond to a 
non-supersymmetric 2HDM~II. 
 
At the end of the day we see that even though the maximum FCNC rates for the 
MSSM and the 2HDM~II are both of order $10^{-4}$ -- and therefore 
potentially visible -- at some point on the road it should be possible to 
disentangle the nature of the Higgs model behind the FCNC decays of the top 
quark. Needless to say, if all the recent fuss at CERN~\cite{TJunk} about 
the possible detection of a Higgs boson would eventually be confirmed in
the future (e.g. by the LHC), this 
could still be interpreted as the discovery of one neutral member of an 
extended Higgs model. 
 
We emphasize our most essential conclusions in a nutshell: i) Detection of 
FCNC top quark decay channels into a neutral Higgs boson would be a blazing 
signal of physics beyond the SM; ii) There is a real chance for seeing rare 
events of that sort both in generic Type~II 2HDM's and in the MSSM. The 
maximum rates for the leading FCNC processes~(\ref{Higgschannels}) and 
$t\rightarrow c\;g$
in the 2HDM~II (resp. in the MSSM) satisfy the relations
\begin{equation} 
BR(t\rightarrow g\,c)<10^{-6}(10^{-5})<BR(t\rightarrow h\,c)\sim10^{-4}\,, 
\label{summary} 
\end{equation} 
where it is understood that $h$ is $h^{0}$ or $H^{0}$, but not both, in the 
2HDM~II; whereas $h$ is most likely $h^{0}$, but it could also be $H^{0}$ 
and $A^{0}$, in the MSSM ; iii) Detection of more than one Higgs channel 
would greatly help to unravel the type of underlying Higgs model. 
 
The pathway to seeing new physics through FCNC decays of the top quark is 
thus potentially open. It is now an experimental challenge to accomplish 
this program using the high luminosity super-colliders round the corner. 
 
\Acknowledgments
One of the authors (J.S) would 
like to thank H.~Haber and T.~Junk for helpful discussions. J.G. thanks
S. Pe{\~n}aranda for suggestions.

\providecommand{\href}[2]{#2}\begingroup\raggedright\endgroup


\begin{thebibliography}{10}

\bibitem{Eilam:1991zc}
G.~Eilam, J.~L. Hewett,  A.~Soni, {\em Phys. Rev.} {\bf D44} (1991)
1473--1484.

\bibitem{Mele:1998ag}
B.~Mele, S.~Petrarca,  A.~Soddu, {\em Phys. Lett.} {\bf B435} (1998) 401,
\href{http://xxx.lanl.gov/abs/hep-ph/9805498}{{\tt hep-ph/9805498}};
G.~Eilam, J.~L. Hewett,  A.~Soni, {\em Phys. Rev.} {\bf D59} (1999)
039901, Erratum.

\bibitem{Lopez:1997xv}
J.~L. Lopez, D.~V. Nanopoulos,  R.~Rangarajan, {\em Phys. Rev.} {\bf D56}
  (1997) 3100--3106,
\href{http://xxx.lanl.gov/abs/hep-ph/9702350}{{\tt hep-ph/9702350}}.

\bibitem{deDivitiis:1997sh}
G.~M. de~Divitiis, R.~Petronzio,  L.~Silvestrini, {\em Nucl. Phys.} {\bf B504}
  (1997) 45--60,
\href{http://xxx.lanl.gov/abs/hep-ph/9704244}{{\tt hep-ph/9704244}}.

\bibitem{Guasch:1999jp}
J.~Guasch,  J.~Sol{\`a}, {\em Nucl. Phys.} {\bf B562} (1999) 3--28,
\href{http://xxx.lanl.gov/abs/hep-ph/9906268}{{\tt hep-ph/9906268}}.

\bibitem{Bejar:2000ub}
S.~B{\'e}jar, J.~Guasch,  J.~Sol{\`a},
\href{http://xxx.lanl.gov/abs/hep-ph/0011091}{{\tt hep-ph/0011091}}.

\bibitem{TopPhys}
M.~Beneke, I.~Efthymiopoulos, M.L.~Mangano, J.~Womersley (conveners),
\textit{Top Quark Physics}, proc. of the workshop \textit{Standard Model physics (and
more) at the LHC}, p. 419, eds. G.~Altarelli and M.L.~Mangano,
\href{http://xxx.lanl.gov/abs/hep-ph/0003033}{{\tt hep-ph/0003033}}.

\bibitem{Duncan:1983iq}
M.~J. Duncan, {\em Nucl. Phys.} {\bf B221} (1983)
285;
{\em Phys. Rev.} {\bf D31} (1985)
1139.

\bibitem{Gabbiani:1996hi}
F.~Gabbiani {\em et al.},
{\em Nucl. Phys.} {\bf B477} (1996) 321--352,
\href{http://xxx.lanl.gov/abs/hep-ph/9604387}{{\tt hep-ph/9604387}};
M.~Misiak, S.~Pokorski,  J.~Rosiek, \textit{Supersymmetry
and FCNC Effects}, 
in: ``Heavy Flavours II'', p. 795, eds. A.J.~Buras, M. Lindner, Advanced Series
  on directions in High Energy Physics, World Scientific 1998,
  \href{http://xxx.lanl.gov/abs/hep-ph/9703442}{{\tt hep-ph/9703442}}.

\bibitem{Hunter}
J.~Gunion, H.~Haber, G.~Kane, S.~Dawson,
{\it{The Higgs Hunter's Guide}} (Addison-Wesley, Menlo-Park, 1990).

\bibitem{Dabelstein}
A.~Dabelstein, {\em Z. Phys.} {\bf C67} (1995) 495--512,
\href{http://xxx.lanl.gov/abs/hep-ph/9409375}{{\tt hep-ph/9409375}};
{\em Nucl. Phys.} {\bf B456} (1995) 25--56,
\href{http://xxx.lanl.gov/abs/hep-ph/9503443}{{\tt hep-ph/9503443}}.

\bibitem{Carena:1996wu}
M.~Carena {\em et al.},
  {\em Nucl. Phys.} {\bf B580} (2000) 29--57,
\href{http://xxx.lanl.gov/abs/hep-ph/0001002}{{\tt hep-ph/0001002}};
J.~R.~Espinosa,  R.-J.~Zhang, {\em Nucl. Phys.} {\bf B586} (2000) 3--38,
\href{http://xxx.lanl.gov/abs/hep-ph/0003246}{{\tt hep-ph/0003246}},
and references therein.

\bibitem{Coarasa:1996qa}
J.~Guasch, R.~A. Jim{\'e}nez,  J.~Sol{\`a}, {\em Phys. Lett.} {\bf B360} (1995)
  47--56,
\href{http://xxx.lanl.gov/abs/hep-ph/9507461}{{\tt hep-ph/9507461}};
J.~A. Coarasa {\em et al.}, 
{\em Eur. Phys. J.} {\bf C2} (1998) 373,
\href{http://xxx.lanl.gov/abs/hep-ph/9607485}{{\tt hep-ph/9607485}}.

\bibitem{Alam:1995aw}
M.~S. Alam {\em et al.},  CLEO Collaboration, {\em Phys. Rev. Lett.} {\bf 74}
  (1995)
2885--2889;
 CLEO Collaboration, proc. of the \textit{29th International Conference on
  High-Energy Physics (ICHEP 98)}, Vancouver, B.C., Canada, 1998;
S.~Ahmed {\em et al.},  CLEO Collaboration,
\href{http://xxx.lanl.gov/abs/hep-ex/9908022}{{\tt hep-ex/9908022}}.

\bibitem{Borzumati:1998tg}
F.~M. Borzumati,  C.~Greub, {\em Phys. Rev.} {\bf D58} (1998) 074004,
\href{http://xxx.lanl.gov/abs/hep-ph/9802391}{{\tt hep-ph/9802391}};
\textit{ibid.} {\bf D59} (1999) 057501,
\href{http://xxx.lanl.gov/abs/hep-ph/9809438}{{\tt hep-ph/9809438}};
\href{http://xxx.lanl.gov/abs/hep-ph/9810240}
Proc. of ICHEP 98,
Vancouver, Canada, 23-29 Jul 1998, vol. 2, 1735-1739; {{\tt
  hep-ph/9810240}}.
\bibitem{Chankowski:1999ta}
P.~H. Chankowski, M.~Krawczyk,  J.~Zochowski, {\em Eur. Phys. J.} {\bf C11}
  (1999) 661, \href{http://xxx.lanl.gov/abs/hep-ph/9905436}{{\tt
  hep-ph/9905436}}, and references therein.

\bibitem{Ciuchini:1998xy}
M.~Ciuchini {\em et al.}, 
{\em Nucl. Phys.} {\bf B534} (1998) 3,
\href{http://xxx.lanl.gov/abs/hep-ph/9806308}{{\tt hep-ph/9806308}};
M.~Carena {\em et al.}, 
\href{http://xxx.lanl.gov/abs/hep-ph/0010003}{{\tt hep-ph/0010003}};
D.~Garcia,  these proceedings,
\href{http://xxx.lanl.gov/abs/hep-ph/0101206}{{\tt hep-ph/0101206}}. 


\bibitem{PDB2000}
D.~E. Groom {\em et al.}, {\em Eur. Phys. J.} {\bf C15} (2000)
1.

\bibitem{Frey:1997sg}
R.~Frey {\em et al.}, FERMILAB-CONF-97-085, April 1997,
\href{http://xxx.lanl.gov/abs/hep-ph/9704243}{{\tt  hep-ph/9704243}}.

\bibitem{Aguilar-Saavedra:2000aj}
J.~A. Aguilar-Saavedra,  G.~C. Branco, {\em Phys. Lett.} {\bf B495} (2000)
  347--356,
\href{http://xxx.lanl.gov/abs/hep-ph/0004190}{{\tt hep-ph/0004190}};
J.~A. Aguilar-Saavedra,
\href{http://xxx.lanl.gov/abs/hep-ph/0012305}{{\tt hep-ph/0012305}}.

\bibitem{TJunk}
T.~Junk, these proceedings, 
\href{http://xxx.lanl.gov/abs/hep-ex/0101015}{{\tt hep-ex/0101015}}, 

\bibitem{Hagiwara}
K.~Hagiwara, these proceedings. 

\end{thebibliography}
\end{document}